\documentclass[pre, showpacs, twocolumn]{revtex4}

\usepackage[latin1]{inputenc}
\usepackage[T1]{fontenc}
\usepackage[british]{babel}

\usepackage{psfrag}
\usepackage{graphicx}
\usepackage{braket}

\usepackage[intlimits]{empheq}
\usepackage{amssymb}
\usepackage{amsthm}
\usepackage{mathtools}
\usepackage{amsfonts}
\usepackage{xspace}
\usepackage{bm}
\usepackage{nicefrac}

\allowdisplaybreaks[1]

\usepackage{hyperref}

\newcommand{\vc}[1]{\ensuremath{\bm{#1}}\xspace}

\newcommand{\Axy}[2]{\hat{A}_{#1}^{\ #2}}
\newcommand{\Anm}{\hat{A}_{n}^{\ m}}

\newcommand{\Ann}{\hat{A}_{n}^{\ n}}

\newcommand{\Azxy}[3]{\hat{a}_{#1, #2}^{(#3)}}

\newcommand{\aop}[1]{\hat{a}_{#1}}
\newcommand{\aopd}[1]{\hat{a}^{\dag}_{#1}}

\newcommand{\ie}{i.\,e.\xspace}

\begin{document}

\title{Thermodynamics and Superradiant Phase Transitions in a
  three-level Dicke Model}
\author{Mathias Hayn}
\author{Tobias Brandes}
\affiliation{Institut f\"ur Theoretische Physik, Technische
  Universit\"at Berlin, 10623 Berlin, Germany}
\date{\today}

\begin{abstract}
	We analyse the thermodynamic properties of a generalised Dicke
  model, \ie a collection of three-level systems interacting with two
  bosonic modes. We show that at finite temperatures the system
  undergoes first-order phase transitions only, which is in contrast
  to the zero-temperature case where a second-order phase transition
  exist as well. We discuss the free energy and prominent expectation
  values. The limit of vanishing temperature is discussed as well.
\end{abstract}


\maketitle

\section{Introduction}
%
The Dicke superradiance model \cite{Dicke1954} as a test-bed for
mean-field like phase transitions \cite{Hepp1973a} has received
renewed attention recently, in particular due to its successful
implementation in cold-atom experiments \cite{Baumann2010,
  Baumann2011, Ritsch2013, Hamner2014, Klinder2015} and optical setups
of cavities and lasers \cite{Dimer2007, Baden2014}. Much progress has
been made towards a realistic description of non-equilibrium and
dissipation \cite{Keeling2010, Nagy2011, Oeztop2012, Bhaseen2012,
  Konya2012, Torre2013, Kopylov2013, Genway2014, DallaTorre2016,
  Gelhausen2016}, multi-mode effects \cite{Konya2011}, the interplay
of the superradiant phase transitions and Bose--Einstein Condensation
\cite{Piazza2013}, spin glasses \cite{Strack2011}, the analysis of
interactions \cite{Inoue2012}, inhomogeneous couplings \cite{Goto2008,
  Tsyplyatyev2009}, finite size effects \cite{Vidal2006}, or the
adiabatic limit \cite{Liberti2006, Bakemeier2012}. Further extensions
of the model have appeared allowing for driving \cite{Bastidas2012,
  Francica2016}, the creation of Goldstone modes \cite{Brandes2013,
  Yi-Xiang2013, Baksic2014}, feedback control \cite{Grimsmo2014,
  Kopylov2015}, or the transfer to other platforms such as solid-state
systems \cite{Nataf2010, Nataf2010a}.

In its simplest version, the Dicke model hosts a mean-field type
ground state phase transition (`quantum bifurcation') at zero
temperature \cite{Hepp1973a, Emary2003, Emary2003a} between a
field-free (normal) phase with unpolarised atoms, and a superradiant
phase with macroscopic occupation of the field mode and polarisation
of the atoms. Above a critical coupling strength, the superradiant
phase also persists at finite temperatures below a critical
temperature which defines the corresponding thermal second order phase
transition \cite{Wang1973, Hepp1973b, Carmichael1973, Brandes2005}.

Thermal aspects of the Dicke-Hepp-Lieb superradiance phase transition
have found recent interest again in the analysis of thermodynamic
aspects like work extraction \cite{Paraan2009, Fusco2016}, and the
discussion of van-Hove type singularities in the microcanonical
density of states at large excitation energies
\cite{Perez-Fernandez2011a, Perez-Fernandez2011b, Brandes2013,
  Puebla2013, Puebla2013a, Bastarrachea-Magnani2014,
  Bastarrachea-Magnani2015, Lobez2016, Bastarrachea-Magnani2016,
  Kloc2016} (excited state quantum phase transitions). This is our
motivation to extend our previous studies \cite{Hayn2011, Hayn2012} of
an extended Dicke model towards finite temperatures in this paper.

In the original Dicke model \cite{Dicke1954}, the atoms are
approximated by two-level systems and the light field by a single
(bosonic) mode of a resonator. Naturally, the questions arises what
happens to the phases and phase transitions when the Dicke model is
generalised by having more than just two (atomic) energy levels and
one bosonic mode. In our previous study at zero temperature, we found
an additional superradiant phase as well as phase transitions of first
and second order in a three-level Dicke model interacting with two
bosonic modes in Lambda-configuration. These finding where also of
interest in the context of the discussion of the no-go theorem
\cite{Rzazewski1975, Rzazewski1976, Rzazewski1976a, Knight1978,
  Bialynicki-Birula1979, Slyusarev1979} for superradiant phase
transitions in this and other generalised Dicke models
\cite{Nataf2010, Viehmann2011, Ciuti2012, Viehmann2012, Hayn2012,
  Baksic2013}.

Our paper is organised as follows: first, we give a short review of
the model and previous findings for zero temperature
\cite{Hayn2011}. Then, the partition sum of the model is calculated in
the thermodynamic limit from which we identify the thermodynamic
phases relevant expectation values. We discuss the properties of the
phases and the phase transitions and finally recover the
zero-temperature results as a limiting case of our theory.

\section{Finite-Temperature Phase Transition in the Lambda-Model}

\subsection{The Model}
\begin{figure}[t]
	\includegraphics[width=6.0cm]{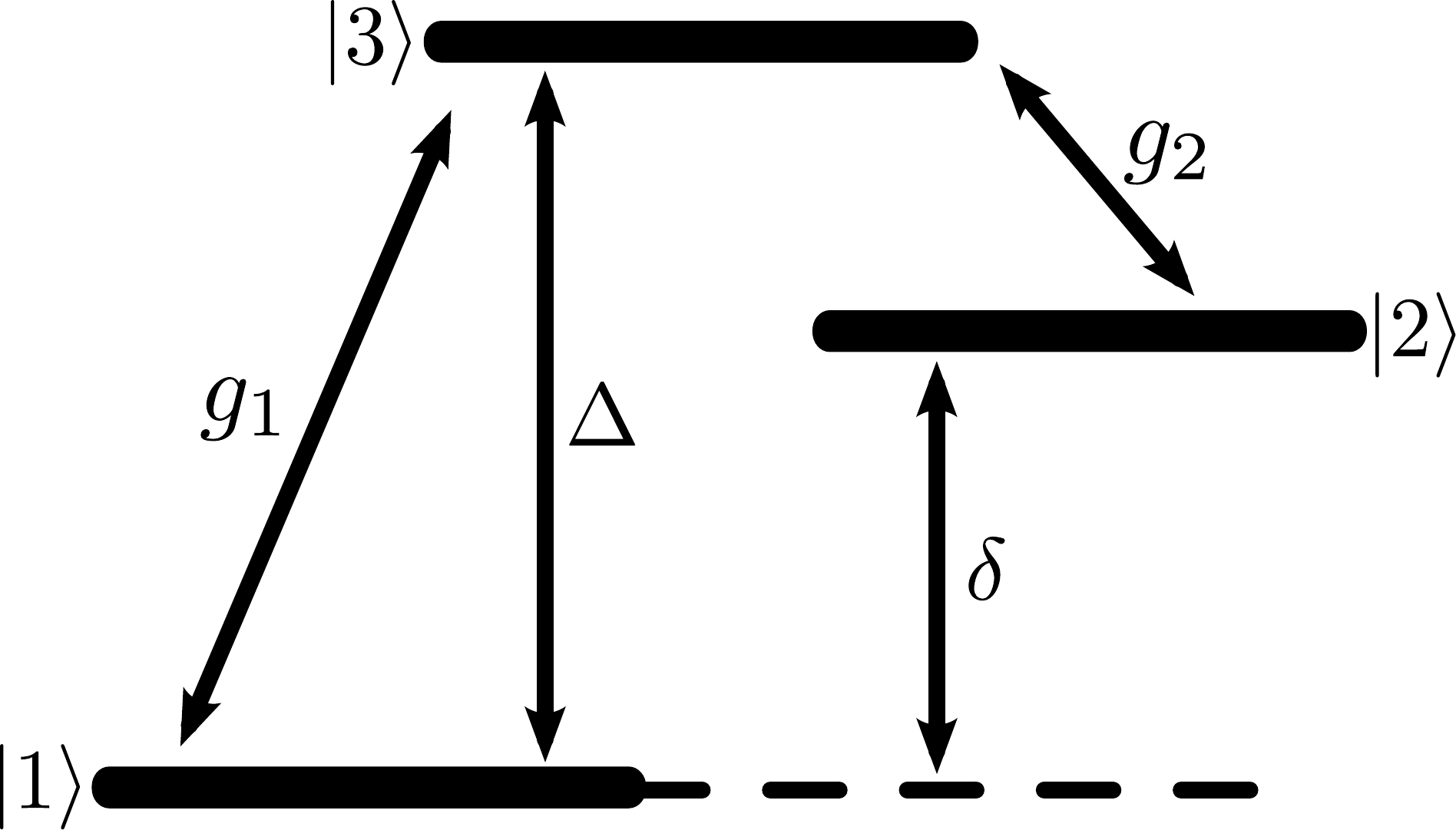}
	\caption{Single-particle energy levels in Lambda-configuration: The
    excited state $\ket 3$ is coupled via the two bosonic
    modes with frequencies $\omega_1$ and $\omega_2$ to either of the
    two ground-states $\ket 1$ and $\ket 2$, respectively. The
    corresponding coupling strengths are given by $g_1$ and
    $g_2$.} \label{fig:model}
\end{figure}

We consider a system of $\mathcal N$ three-level systems (the
\textsl{particles}) in Lambda-configuration
(cf. Fig.~\ref{fig:model}): two ground-states $\ket 1$, $\ket 2$ with
energies $E_1$, $E_2$, respectively, are coupled via two bosonic modes
to the excited state $\ket 3$ with energy $E_3$. The two bosonic modes
have frequencies $\omega_1$, $\omega_2$, respectively. The Hamiltonian
is given by (cf. Ref.~\cite{Hayn2011})
\begin{multline} \label{eq:HamLambda99}
  \hat H = \delta \Axy{2}{2} + \Delta \Axy{3}{3} + \hbar \omega_1
  \aopd{1} \aop{1} + \hbar \omega_2 \aopd{2} \aop{2} \\
  + \frac{g_1}{\sqrt{\mathcal N}} \bigl( \aopd{1} + \aop{1} \bigr)
  \bigl( \Axy{1}{3} + \Axy{3}{1} \bigr) + \frac{g_2}{\sqrt{\mathcal
      N}} \bigl( \aopd{2} + \aop{2} \bigr) \bigl( \Axy{2}{3} +
  \Axy{3}{2} \bigr).
\end{multline}
Here, $\delta = E_2 - E_1$, $\Delta = E_3 - E_1$ and $g_n$ is the
coupling strength of $n^{\text{th}}$ bosonic mode. The operators
$\Anm, n,m \in \{1,2,3\}$ are collective particle operators and can be
written in terms of single-particle operators $\Azxy n m k$,
\begin{equation}
  \Anm = \sum_{k=1}^{\mathcal N} \Azxy n m k.
\end{equation}
The operator $\Azxy n m k$ acts on the degrees of freedom of the
$k^{\text{th}}$ particle only and can be represented by $\Azxy n m k =
\ket n^{(k)} \bra m$, where $\ket n^{(k)}$ is the $n^{\mathrm{th}}$
state of the $k^{\mathrm{th}}$ three-level system.

We call the single-particle energy levels $\ket 1$ ($\ket 2$) and $\ket 3$
together with the first (second) bosonic mode, the left (right) branch
of the Lambda-model.

In a previous work~\cite{Hayn2011}, we have studied the ground-state
properties of the Hamiltonian, Eq.~\eqref{eq:HamLambda99}, as well as
collective excitations above the ground-state. The model shows three
phases: a normal phase and two superradiant phases. Both types of
phases show their distinctive features as in the original Dicke model
\cite{Emary2003}. The normal phase has zero occupation of the bosonic
modes and all three-level systems occupy their respective
single-particle ground-state $\ket 1$. In contrast, the superradiant
phases are characterised by a macroscopic occupation of the bosonic
modes and the three-level systems. The superradiant phases are
divided into a so-called blue and red superradiant phase. For the
blue superradiant phase, the left branch of the Lambda-model is
macroscopically occupied, whereas for the red superradiant phase it is
the right branch which shows macroscopic occupation. The phase diagram
features phase transitions of different order; the phase transition
between the blue superradiant phase and the normal phase is
continuous. In contrast, for finite $\delta$, the phase transition
between the normal and the red superradiant phase is of first
order. Eventually, in the limit $\delta \rightarrow 0$ this phase
transition becomes continuous as well. The two superradiant phases are
separated by a first order phase transition, irrespective of the
parameters of the model.

\subsection{Evaluation of the Partition Sum}
All thermodynamic information of the equilibrium system is contained
in the partition sum \cite{Feynman1972}, $\mathcal Z = \mathcal
Z(\mathcal N, T) = \text{Tr} \{ \mathrm \exp [- \beta \hat H]
\}$. Here, $T$ is the temperature and $\beta = 1/(k_{\mathrm B} T)$,
with Boltzmann's constant $k_{\mathrm B}$.

In order to evaluate the trace, we represent the bosonic degrees of
freedom by coherent states \cite{Glauber1963, Arecchi1972} with
respect to the $\hat a_n$ and the trace of the particle degrees of
freedom are split into single-particle traces $\mathrm{Tr}_n$,
\begin{equation}
  \mathcal Z = \int_{\mathbb C} \frac{\mathrm d^2 \alpha_1}{\pi}
  \int_{\mathbb C} \frac{\mathrm d^2 \alpha_2}{\pi} \braket{\alpha_1,
    \alpha_2 | \Bigl( \prod_{n=1}^{\mathcal N} \mathrm{Tr}_n \Bigr)
    \mathrm e^{-\beta \hat H} | \alpha_1, \alpha_2}.
\end{equation}
Both integrals extend over the complex plane, respectively.  Due to
the coherent states, the bosonic part of the trace is easily
computed. For the particle part of the trace, we observe that the
$\mathcal N$ single-particle traces are all identical. In addition, we
decompose both $\alpha_1$ and $\alpha_2$ in its real and imaginary
part and scale them with $\sqrt{\mathcal N}$,
\begin{equation}
  \alpha_n = \sqrt{\mathcal N \,} y_n + i \sqrt{\mathcal N \,} z_n, \;
  n = 1,2.
\end{equation}
Considering the thermodynamic limit $\mathcal N \rightarrow \infty$,
the partition sum eventually reads
\begin{equation} \label{eq:LMPartSum1} \mathcal Z = \frac{\mathcal
    N^2}{\pi^2} \int_{\mathbb R^4} \mathrm dy_1 \mathrm dz_1 \mathrm
  dy_2 \mathrm dz_2 \, \mathrm e^{-\beta \mathcal N \sum_{n=1}^2 \hbar
    \omega_n (y_n^2 + z_n^2)} \mathrm{Tr} \Bigl\{ \mathrm
  e^{-\beta \hat h} \Bigr\}^{\mathcal N}.
\end{equation}
Here, the single-particle Hamiltonian $\hat h = \hat h (y_1, y_2)$ is
given by
\begin{multline} \label{eq:LM3Ham}
  \hat h(y_1, y_2) = \delta \hat a_{2,2} + \Delta \hat a_{3,3} \\
  + 2 g_1 y_1 \bigl( \hat a_{1,3} + \hat a_{3,1} \bigr) + 2 g_2 y_2
  \bigl( \hat a_{2,3} + \hat a_{3,2} \bigr).
\end{multline}
Since all particles are identical, we have omitted the superindex
$k$ at the single-particle operators $\Azxy n m k$.

The remaining single-particle trace is evaluated in the eigenbasis of
the single-particle Hamiltonian,
\begin{equation} \label{eq:LM3Matrix}
  \hat h = 
  \begin{pmatrix}
    0 & 0 & 2 g_1 y_1 \\
    0 & \delta & 2 g_2 y_2 \\
    2 g_1 y_1 & 2 g_2 y_2 & \Delta
  \end{pmatrix},
\end{equation}
where we have chosen a convenient basis for the matrix representation
of $\hat a_{n,m}$.

By virtue of Cardano's formula, the eigenvalues of $\hat h$ can be
calculated exactly.  However, the discussion of whether there is a
phase transition or not and the analysis of the phase transition, is
not very transparent. Therefore, we will pass the general case to a
numerical computation and first consider the special case with $\delta
= 0$ only, which is amenable to analytical calculations.

For $\delta = 0$, the partition sum can be written in compact form as
(again in the thermodynamic limit $\mathcal N \rightarrow \infty$)
\begin{equation} \label{eq:LMPartSum2}
  \mathcal Z = \frac{\mathcal N^2}{\pi^2} \int_{\mathbb R^2} \mathrm
  dy_1 \mathrm dz_1 \int_{\mathbb R^2} \mathrm dy_2 \mathrm dz_2 \,
  \mathrm e^{-\mathcal N f},
\end{equation}
with
\begin{multline} \label{eq:LM_Def_fy}
  f = f(y_1, y_2, z_1, z_2) = \beta \hbar \omega_1 y_1^2 + \beta \hbar
  \omega_2 y_2^2 + \beta \hbar \omega_1 z_1^2 \\ + \beta \hbar \omega_2
  z_2^2 - \ln \Bigl[ 1 + 2 \mathrm e^{-\beta \Delta / 2} \, \cosh \Bigl(
    \frac{\beta \Delta}{2} \Omega \Bigr) \Bigr]
\end{multline}
and a corresponding $\Omega$ given by
\begin{equation}
  \Omega = \Omega(y_1, y_2) = \sqrt{ 1 + 16 g_1^2 y_1^2 / \Delta^2 +
    16 g_2^2 y_2^2 / \Delta^2 \,}.
\end{equation}
The remaining integrals cannot be done exactly. However, we can
approximate them for large $\mathcal N$ by Laplace's
method~\cite{Bender1999}. Since large values of $f$ are exponential
suppressed, the main contribution to the integral is given by the
global minimum of $f$. Given the minimum, the partition sum is
proportional to
\begin{equation}
  \mathcal Z \propto \mathrm  e^{-\mathcal N f_0},
\end{equation}
where $f_0 = f(y_{1,0}, y_{2,0})$ with the position the minimum
$(y_{1,0}, y_{2,0})$. We anticipate that both $z_n$ are identical zero
at the minimum. This will be shown below. Hence, the leading
contribution to the free energy $F$ \cite{Feynman1972}
is given by $F = \mathcal N k_{\mathrm B} T f_0$.

Before we determine the minimum of $f$, we will first compute
expectation values of observables using the same approximations as
above.

\subsection{Expectation Values}

In order to identify the phases and phase transitions, we discuss
several observables. Of interest are the occupations of the bosonic
modes, $\braket{\hat a^\dag_m \hat a_m}, (m = 1,2)$, and the
three-level systems, $\braket{\Ann}, (n = 1,2,3)$. In addition, the
quantities $\braket{\Axy 1 3}$ and $\braket{\Axy 2 3}$ are
considered. The real part of these give the macroscopic polarisations
of the three-level systems of the left and the right branch of the
Lambda-model, respectively. They are generalisations of the
polarisation in the Dicke model. There, the polarisation is
proportional to the expectation value of the $x$ component of the
atomic pseudo spin operator.

For functions $G$ of operators of the two bosonic modes, the
expectation value is given by (the calculation can be found in the
appendix~\ref{sec:AppA})
\begin{multline} \label{eq:LM_ExpValModeOp}
  \braket{G(\hat a^\dag_1, \hat a_1, \hat a^\dag_2, \hat a_2)} \\
  = G(\sqrt{\mathcal N \,} y_{1,0}, \sqrt{\mathcal N \,} y_{1,0},
  \sqrt{\mathcal N \,} y_{2,0}, \sqrt{\mathcal N \,} y_{2,0}).
\end{multline}

With that, we obtain for the occupation of both modes
\begin{equation} \label{eq:OccModes}
  \braket{\hat a^\dag_n \hat a_n} = \mathcal N \, y_{n,0}^2.
\end{equation}

\label{sec:LM_ExpValAtoms}
Expectation values of collective operators of the three-level systems
can be calculated in a similar manner. Let $\hat M$ be a collective
operator and $\hat m^{(n)}$ the corresponding single-particle operator
for the $n^{\mathrm{th}}$ three-level system, such that
\begin{equation}
  \hat M = \sum_{n=1}^{\mathcal N} \hat m^{(n)}.
\end{equation}
Then, the mean value of $\hat M$ is given by (the calculation can be
found in the appendix~\ref{sec:AppB})
\begin{equation}
  \braket{\hat M} = \mathcal N \braket{\hat m}_0.
\end{equation}
Here, $\braket{\cdot}_0$ is the single-particle expectation value for
a thermal state with the single-particle Hamiltonian $\hat h$
evaluated at the minimum $(y_{1,0}, y_{2,0})$ of $f$. Thus, the
expectation values for the collective atomic operators $\Anm$ can be
traced back to the single-particle operators $\hat a_{n,m}$,
\begin{equation} \label{eq:LM_ExpValAnm1}
  \braket{\Anm} = \mathcal N \braket{\hat a_{n,m}}_0.
\end{equation}
This derivation and the following calculation holds even for non-zero
$\delta$.

Let $\varepsilon_n$ be the eigenvalues and $\vc w_n$ the corresponding
eigenvectors of $\hat h_0$. Then we evaluate the traces in
Eq.~\eqref{eq:LM_ExpValAnm1} in this eigenbasis and the expectation
values can be written as
\begin{equation} \label{eq:LM_ExpValAnm2}
  \braket{\Anm} = \mathcal N \frac{1}{z} \sum_{k=1}^3 (\vc
  w_k^*)_n (\vc w_k)_m \, \mathrm e^{-\beta \varepsilon_k}.
\end{equation}
Here $z = \mathrm{Tr} \{ \exp[-\beta \hat h_0] \}$ is the partition
sum of the single-particle Hamiltonian $\hat h_0$ and we have used
that the matrix elements of $\hat a_{n,m}$ are given by zeros, except
for the entry of the $n^{\mathrm{th}}$ row and $m^{\mathrm{th}}$
column which is one.

\subsection{Partition Sum}
We continue with the calculation of the partition sum,
Eq.~\eqref{eq:LMPartSum2}. Therefore, we need to find the minimum of
$f$, Eq.~\eqref{eq:LM_Def_fy}. This analysis will be done in the
following.

The variable $z_n$ enters only quadratically in $f$, such that upon
minimising $f$, both $z_n$ need to be zero. Minimising $f$ with
respect to $y_n$ yields the two equations ($n=1,2$)
\begin{equation} \label{eq:LMFMinCond}
  0 = y_n \Biggl[ \Bigl( \frac{g_{n,c}}{g_n^2} \Bigr)^2 \Omega -
  q(\Omega) \Biggr],
\end{equation}
with
\begin{equation} \label{eq:LM_Def_gOmega}
  q(\Omega) = \frac{2 \mathrm e^{-\beta \Delta / 2} \sinh \bigl[ \frac{\beta
  \Delta}{2} \Omega \bigr]}{1 + 2 \mathrm e^{-\beta  \Delta
    / 2} \cosh \bigl[ \frac{\beta \Delta}{2} \Omega \bigr]},
\end{equation}
and
\begin{equation}
  g_{n,c} = \frac{\sqrt{\Delta \hbar \omega_n \,}}{2}.
\end{equation}

Of course, Eqs.~\eqref{eq:LMFMinCond} are always solved by the trivial
solutions $y_1 = y_2 = 0$. But do non-trivial solutions exist, and for
which parameter values?

We first observe that the Eqs.~\eqref{eq:LMFMinCond} do not support
solutions where both $y_1$ and $y_2$ are non-zero. For given $y_1$ and
$y_2$, the parameter $\Omega = \Omega(y_1, y_2)$ is fixed. Then the
squared bracket cannot be zero for both equations. Hence, the
non-trivial solutions are given by one $y_n$ being zero and the other
being finite. In the following, the non-zero solution will be called
$y_{n,0}$.

To check whether a non-zero $y_{n,0}$ really exists, we have to analyse the
equation
\begin{equation} \label{eq:LMFMinCondn}
  0 = \Bigl( \frac{g_{n,c}}{g_n} \Bigr)^2 \Omega_{n,0} - q \bigl(
  \Omega_{n,0} \bigr),
\end{equation}
with
\begin{equation} \label{eq:LM_Def_Omega3}
  \Omega_{n,0} = \sqrt{ 1 + 4 \frac{\hbar \omega_n}{\Delta} \Bigl(
    \frac{g_n}{g_{n,c}} \Bigr)^2 y_{n,0}^2 }.
\end{equation}

\begin{figure}[t]
  \includegraphics[width=8.5cm]{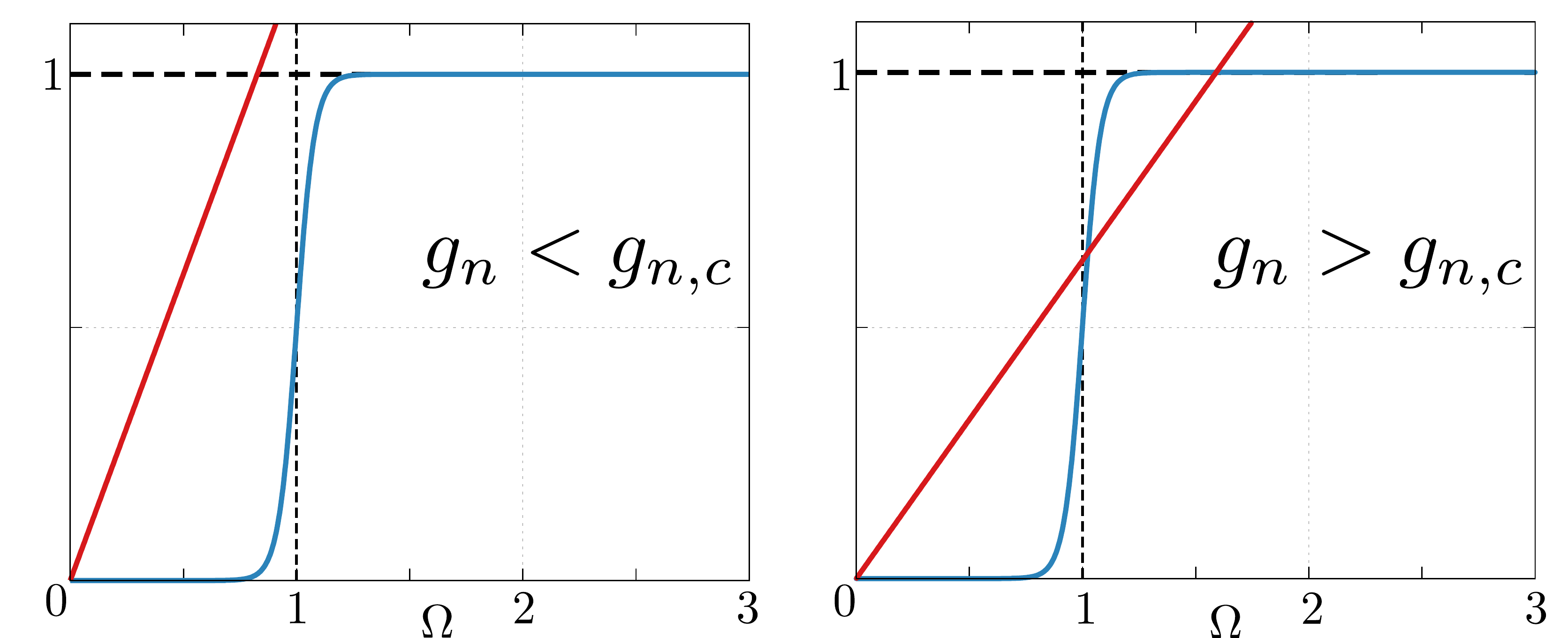}
	\caption{(Color online) Graphical analysis of
    Eq.~\eqref{eq:LMFMinCondn} for $g_n < g_{n,c}$ (left) and $g_n >
    g_{n,c}$ (right) for a certain $n$ and fixed temperature. The red
    straight line corresponds to the first term, $\bigr(
    \frac{g_n}{g_{n,c}} \bigr)^2 \Omega$, the other blue curved line
    corresponds to $q(\Omega)$ of Eq.~\eqref{eq:LMFMinCondn} involving
    the hyperbolic functions. Of physical relevance is the region
    $\Omega > 1$ only. Thus, for $g < g_{n,c}$ no physical solution is
    possible, whereas for $g_n > g_{n,c}$ a physical solution might
    exist.} \label{fig:LMDiscussPT}
\end{figure}

The function $q(\Omega)$ is bounded by one (see left panel of
Fig.~\ref{fig:LMDiscussPT}) and $\Omega$ itself is always greater or
equal one. Therefore, for $g_n < g_{n,c}$, Eq.~\eqref{eq:LMFMinCondn}
has no solution and $y_n$ has to be zero as well.

On the other hand, for $g_n > g_{n,c}$, non-trivial solutions of
Eq.~\eqref{eq:LMFMinCondn} can exist. In the right panel of
Fig.~\ref{fig:LMDiscussPT}, both terms of Eq.~\eqref{eq:LMFMinCondn}
are drawn. We see that for every finite temperature, the two curves
always intersect twice, so that Eq.~\eqref{eq:LMFMinCondn} always has two
solutions. Of course, for a differentiable $f$, the two solutions
cannot both correspond to minima of $f$. Hence, one solution stems
from a maximum and the other from a minimum. Since the right side of
Eq.~\eqref{eq:LMFMinCondn} is the derivative of $f$, its sign-change
signals whether a maximum (plus-minus sign change) or a minimum
(minus-plus sign change) is passed when $\Omega$ is
increased. Therefore, the first solution corresponds to a maximum and
the second to a minimum.

\begin{figure}[t]
  \includegraphics[width=6.0cm]{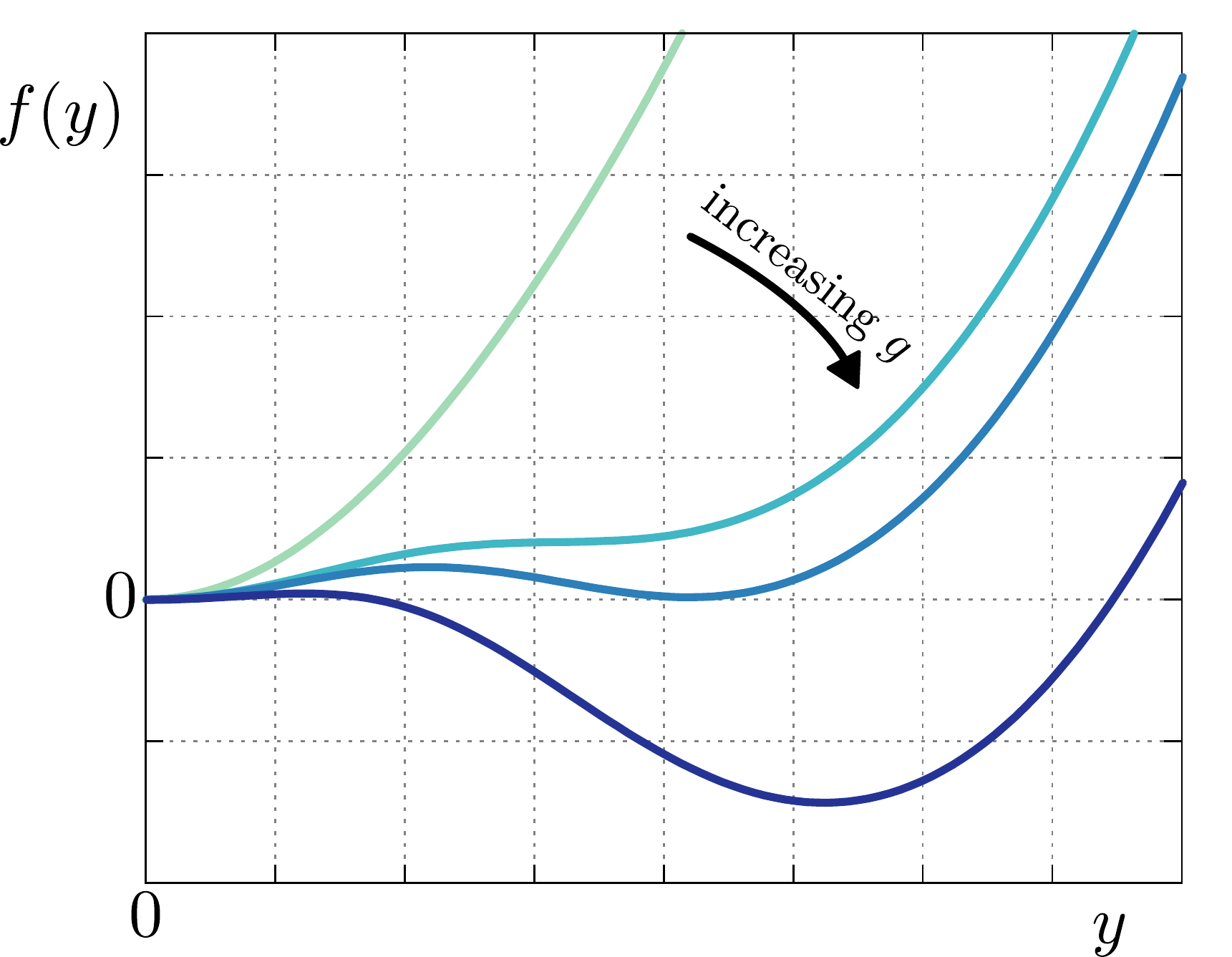}
	\caption{(Color online) The function $f$ of Eq.~\eqref{eq:LM_Def_fy}
    with $z_n = 0 = y_2$, and $y_1 = y$ for fixed temperature. The
    coupling strength $g$ increases from the upper to the lower
    curves. The units are arbitrary and $f$ has been rescaled for
    comparison, such that $f(0) = 0$ for all coupling strengths
    $g$. On increasing $g$, a local minimum forms distant from the
    origin. For large $g$ this minimum eventually becomes a global
    minimum.} \label{fig:LMF_plots}
\end{figure}

We gain additional insight, if we directly analyse $f(y_n)$ for
different coupling strengths [$f(y_n) \equiv f(y_1, 0, 0, 0)$ from
Eq.~\eqref{eq:LM_Def_fy}, w.l.o.g. $n=1$]. This is shown in
Fig.~\ref{fig:LMF_plots}. We see that for small coupling strengths,
$f(y_n)$ has one local minimum only which is located at $y_n=0$, the
trivial solution.
If the coupling strength is increased, a maximum-minimum pair forms at
finite values of $y_n$. In general, this local minimum at $y_n>0$ is
energetically higher than the local minimum of the trivial solution at
$y_n=0$, cf. Fig.~\ref{fig:LMF_plots}. Hence, the trivial solution
still minimises $f(y_n)$ globally. However, if the coupling strength
is increased even further, the local minimum at $y_n > 0$ becomes the
global minimum. So we see that the position $y_{n,0}$ of the global
minimum jumps at a certain value of the coupling strength from zero to
a finite value.

The above discussion refers to the case $\delta = 0$. However, for
finite $\delta$, the results are qualitatively the same. We discuss
the properties of the phase transition for finite $\delta$ below.

\section{The Phase Transition}
In the above analysis, we have shown the existence of three different
minima of $f$ appearing in the exponent of the integrand of the
partition sum. We have also shown that for a given temperature $T$, we
find coupling strengths $g_{1,c}(T), g_{2,c}(T)$, below which the
trivial solution $y_{1,0} = y_{2,0} = 0$ minimises $f$. In
Eq.~\eqref{eq:OccModes}, $y_{n,0}$ measures the macroscopic occupation
of the bosonic modes and can therefore serve as an order parameter of
the phase transition. Hence, in the parameter regime where the trivial
solution minimises $f$, the system is in the normal phase.

In addition, above the coupling strengths $g_{1,c}(T), g_{2,c}(T)$,
$f$ is minimised by non-zero values of either $y_{1,0}$ or
$y_{2,0}$. The first corresponds to the red superradiant phase, the
latter to the blue superradiant phase of Ref.~\cite{Hayn2011}. The two
superradiant phases at non-zero temperatures show only one
macroscopically occupied bosonic mode as well; mode one for the blue
superradiant phase and mode two for the red superradiant phase, while
the occupation of the other mode is zero.

As we have seen in the previous section, the location of the global
minimum $y_{n,0}$ jumps from the trivial solution to non-zero
solutions. A jump of the order parameter defines a first-order
phase transition. Therefore, the finite-temperature superradiant phase
transition in the Lambda-model is a first-order phase
transition. Hence, the continuous phase transition at zero temperature
of Ref.~\cite{Hayn2011} transforms into a first order phase transition
at finite temperatures.

So far, we have identified the phases and phase transitions using the
occupation of the bosonic modes. In the following, we will
further characterise the phases using observables of the three-level
systems.

\subsection{Normal Phase}
\label{sec:NormalPhasee}

In the normal phase with $y_{1,0} = y_{2,0} = 0$, the Hamiltonian
$\hat h_0$ is diagonal and the eigenvectors $\vc w_k$ are given by
Cartesian unit vectors. Hence, the expectation value of all collective
operators $\Anm$ with $n \neq m$ vanish. Conversely for the diagonal
operators $\Axy n n$, the occupations; their expectation values are
given by
\begin{align}
  \braket{\Axy 1 1} &= \mathcal N \, \frac{1}{1 + \mathrm e^{-\beta
      \delta} + \mathrm e^{-\beta \Delta}}, \\
  \braket{\Axy 2 2} &= \mathcal N \, \frac{\mathrm e^{-\beta
    \delta}}{1 + \mathrm e^{-\beta \delta} + \mathrm e^{-\beta \Delta}}, \\
\intertext{and}
  \braket{\Axy 3 3} &= \mathcal N \, \frac{\mathrm e^{-\beta
    \Delta}}{1 + \mathrm e^{-\beta \delta} + \mathrm e^{-\beta \Delta}}.
\end{align}
Here, we explicitly see that in the normal phase the expectation
values are independent of the coupling strengths $g_1$ and
$g_2$. Furthermore we note that for finite temperatures, in addition
to the single-particle ground state $\ket 1$, the energetically higher
lying single-particle states $\ket 2$ and $\ket 3$ are macroscopically
excited as well. Hence in contrast to the bosonic modes, the particle
part of the system gets thermally excited. From that point of view,
\ie concerning the populations of the single-particle energy levels,
the particle system in the normal phase behaves like a \textsl{normal}
thermodynamical system.

\subsection{Superradiant Phases}

For the superradiant phases, we cannot give explicit expressions for
the expectation values, neither for finite or vanishing $\delta$. That
is because we need to compute the minimum of $f$ numerically. Though
for $\delta = 0$, we can say that some expectation values are exactly
zero. This will be done next, separately for the red and the blue
superradiant phases.

\paragraph{Red Superradiant Phase} First consider the red superradiant
phase with $y_{1,0} \equiv y_0 \neq 0$ and $y_{2,0} = 0$. Then, the
single-particle Hamiltonian $\hat h(y_0, 0)$ reads
[cf. Eq.~\eqref{eq:LM3Matrix}]
\begin{equation}
  \hat h (y_0, 0) =
  \begin{pmatrix}
    0 & 0 & 2 g_1 y_0 \\
    0 & 0 & 0 \\
    2 g_1 y_1 & 0 & \Delta
  \end{pmatrix},
\end{equation}
and its exponential has the form
\begin{equation} \label{eq:LMRSRExpOp}
  \mathrm e^{-\beta \hat h (y_0, 0)} =
  \begin{pmatrix}
    a_+ & 0 & b_1 \\
    0 & 1 & 0 \\
    b_1 & 0 & a_-
  \end{pmatrix}.
\end{equation}
The matrix elements $a_\pm$ and $b_n$ are given by (the matrix element
$b_2$ is needed below).
\begin{align}
  a_\pm &= \mathrm e^{-\frac{\beta \Delta}{2}} \left( \cosh \Bigl[
  \frac{\beta \Delta \Omega}{2} \Bigr] \pm \frac{\sinh \Bigl[
    \frac{\beta \Delta \Omega}{2} \Bigr]}{\Omega}
  \right), \label{eq:LM_apm} \\
 b_n &= - \frac{4 g_n y_n}{\Delta \Omega} \mathrm e^{-\frac{\beta
     \Delta}{2}} \, \sinh \Bigl[ \frac{\beta \Delta \Omega}{2}
 \Bigr], \quad n=(1,2). \label{eq:LM_b12}
\end{align}
The product of the exponential operator, Eq.~\eqref{eq:LMRSRExpOp},
with matrices of the form
\begin{equation}
  \begin{pmatrix}
    0 & M_{12} & 0 \\
    M_{21} & 0 & M_{23} \\
    0 & M_{32} & 0
  \end{pmatrix}
\end{equation}
is traceless. Therefore, the expectation values of the collective
operators $\Axy 1 2$, $\Axy 2 3$ and their Hermitian conjugates are
zero, \ie there is no spontaneous polarisation between both the
single-particle states $\ket 1$ and $\ket 2$, and the single-particle
states $\ket 2$ and $\ket 3$. Contrary, the polarisation in the left
branch of the Lambda-model, \ie between the states $\ket 1$ and $\ket
3$, is finite and macroscopic.

\paragraph{Blue Superradiant Phase} For the blue superradiant case,
the discussion is similar. Here we have $y_{2,0} \equiv y_0 \neq 0$
and $y_{1,0} = 0$, and the exponential of the single-particle
Hamiltonian reads
\begin{equation} \label{eq:LMBSRExpOp}
  \mathrm e^{-\beta \hat h(0, y_0)} = 
  \begin{pmatrix}
    1 & 0 & 0 \\
    0 & a_+ & b_2 \\
    0 & b_2 & a_-
  \end{pmatrix}.
\end{equation}
The matrix elements are given above, Eqs~\eqref{eq:LM_apm},
\eqref{eq:LM_b12}.  Now, the product of the exponential operator,
Eq.~\eqref{eq:LMBSRExpOp}, with matrices of the form
\begin{equation}
  \begin{pmatrix}
    0 & M_{12} & M_{13} \\
    M_{21} & 0 & 0 \\
    M_{31} & 0 & 0
  \end{pmatrix}
\end{equation}
is traceless and thus expectation values of the collective operators
$\Axy 1 2$, $\Axy 1 3$ and their Hermitian conjugates are zero. On the
other hand, the expectation value of the operators $\Axy 2 3$, $\Axy 3
2$ is finite and macroscopic. Hence, only the transition in the right
branch of the Lambda-model is spontaneously polarised.

In conclusion, we find that in the superradiant phases at finite
temperature, only the corresponding branch of the Lambda-model shows
spontaneous polarisation; the left branch in the red superradiant
phase and the right branch in the blue superradiant phase. In the
normal phase, the polarisation is completely absent. Hence, in
contrast to the populations of the atomic system, the polarisations
are not thermally excited and show a genuine quantum character. Thus,
both the polarisations and the occupations of the two resonator modes
show a similar behaviour in the three phases. Therefore we have two
sets of observables, the polarisations for the three-level systems and
the occupations of the bosonic modes, to detect the superradiant phase
transition at finite temperatures.

\subsection{Numerical evaluation of the Partition Sum}
The above analysis for vanishing $\delta$ already shows that the phase
transition in the Lambda-model for finite temperatures is a
first-order phase transition. This fact renders the calculation of the
exact location of the phase transition with our methods
impossible. This can be understood with the help of the free energy as
follows. In the thermodynamic limit, the global minimum of the free
energy defines the thermodynamic phase of the system. We explicitly
saw this when we have computed the partition sum. In a phase
transition, the system changes from one thermodynamic state to another
thermodynamic state. This new state corresponds to a different, now
global minimum of the free energy.

For continuous phase transitions, the new minimum evolves continuously
from the first minimum and the first minimum changes its character to a
maximum. Hence, the continuous phase transition is characterised by a
sign-change of the curvature of the free energy at the position of the
minimum of the state describing the normal phase. Often, this is
tractable analytically.

In contrast in the case of first-order phase transitions, the new
global minimum of the free energy appears distant from the old global
minimum of the free energy, cf. Fig.~\ref{fig:LMF_plots}. There are
still two minima and we cannot detect the phase transition by the
curvature of the free energy. Thus to find the phase transition for
first-order phase transitions, we first need to find all minima of the
free energy and then find the global minima of these. This has to be
done numerically here.

\begin{figure}[t]
  \centering
  \includegraphics[width=8.5cm]{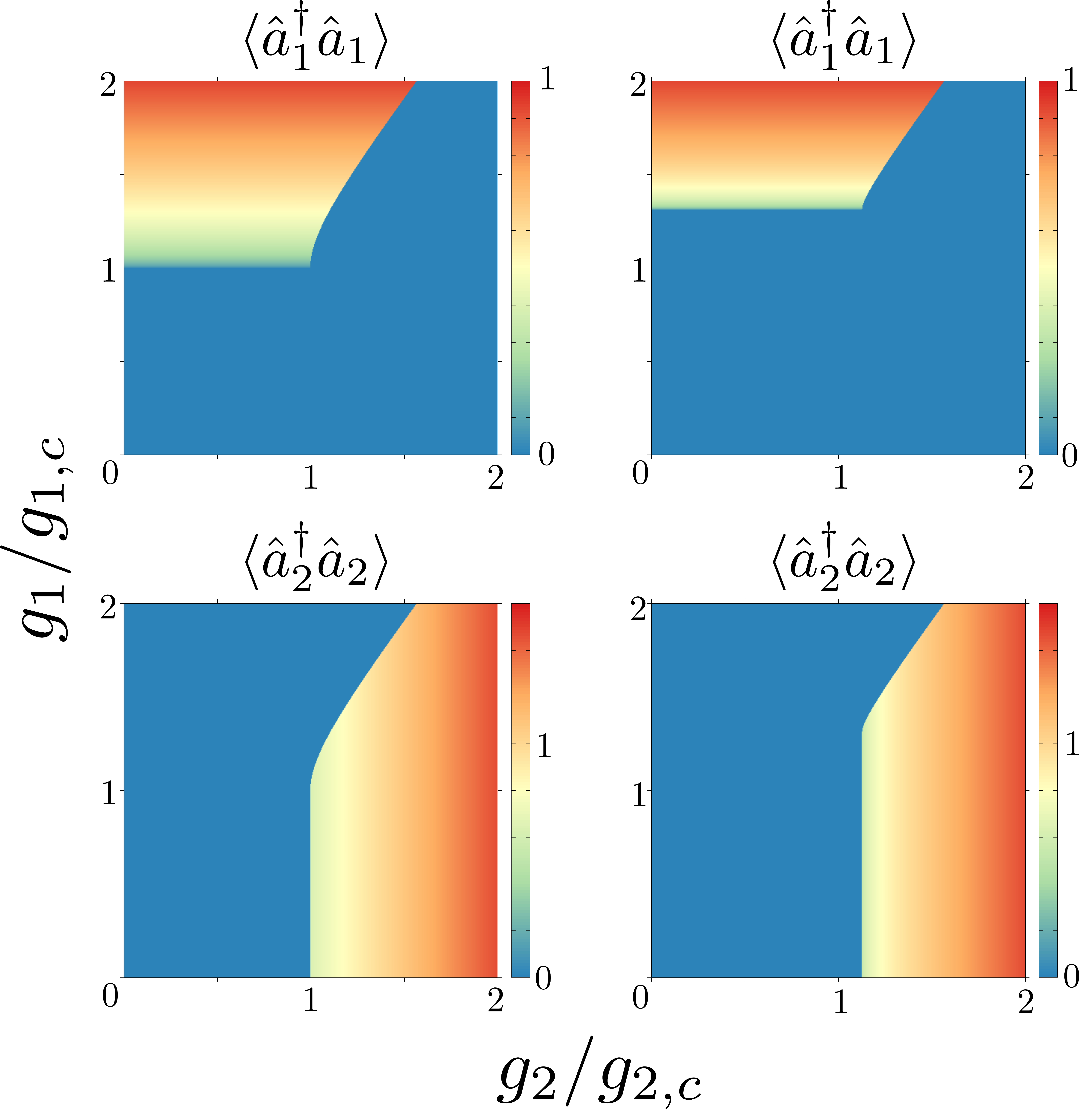}
	\caption{(Color online) Numerical computation of the scaled
    occupations $\braket{\hat a_n^\dag \hat a_n} / \mathcal N$ of the
    two bosonic modes for $k_{\mathrm B} T = 0.001 \, \Delta$ (left) and
    $k_{\mathrm B} T = 0.25 \, \Delta$ (right). The parameters are set to
    $\Delta=1$, $\delta=0.1$, $\omega_1=1.1$, and $\omega_2=0.8$. The
    computation has been done in the thermodynamic limit using
    Laplace's method.} \label{fig:LM_ModeOccTs}
\end{figure}

For the numerical computation, we do not solve
Eq.~\eqref{eq:LMFMinCond}, but we test for the minima of $f(y_1,
y_2)$, Eq.~\eqref{eq:LM_Def_fy}, directly. Therefore, we apply a
brute-force method, \ie we look for the smallest value of $f(y_1,
y_2)$ on a $y_1$--$y_2$ grid. Due to the reflection symmetry of
$f(y_1, y_2)$, we can confine the grid to positive values for $y_1$
and $y_2$. This yields the position of the minimum $(y_{1,0},
y_{2,0})$. Then we compute the eigenvalues and eigenvectors of the
Hamiltonian $\hat h$, at this point and obtain via
Eq.~\eqref{eq:LM_ExpValModeOp} the expectation values for the
operators of the bosonic modes, and via Eq.~\eqref{eq:LM_ExpValAnm2}
the corresponding expectation values for the three-level systems.

\begin{figure}[t]
  \centering
  \includegraphics[width=8.5cm]{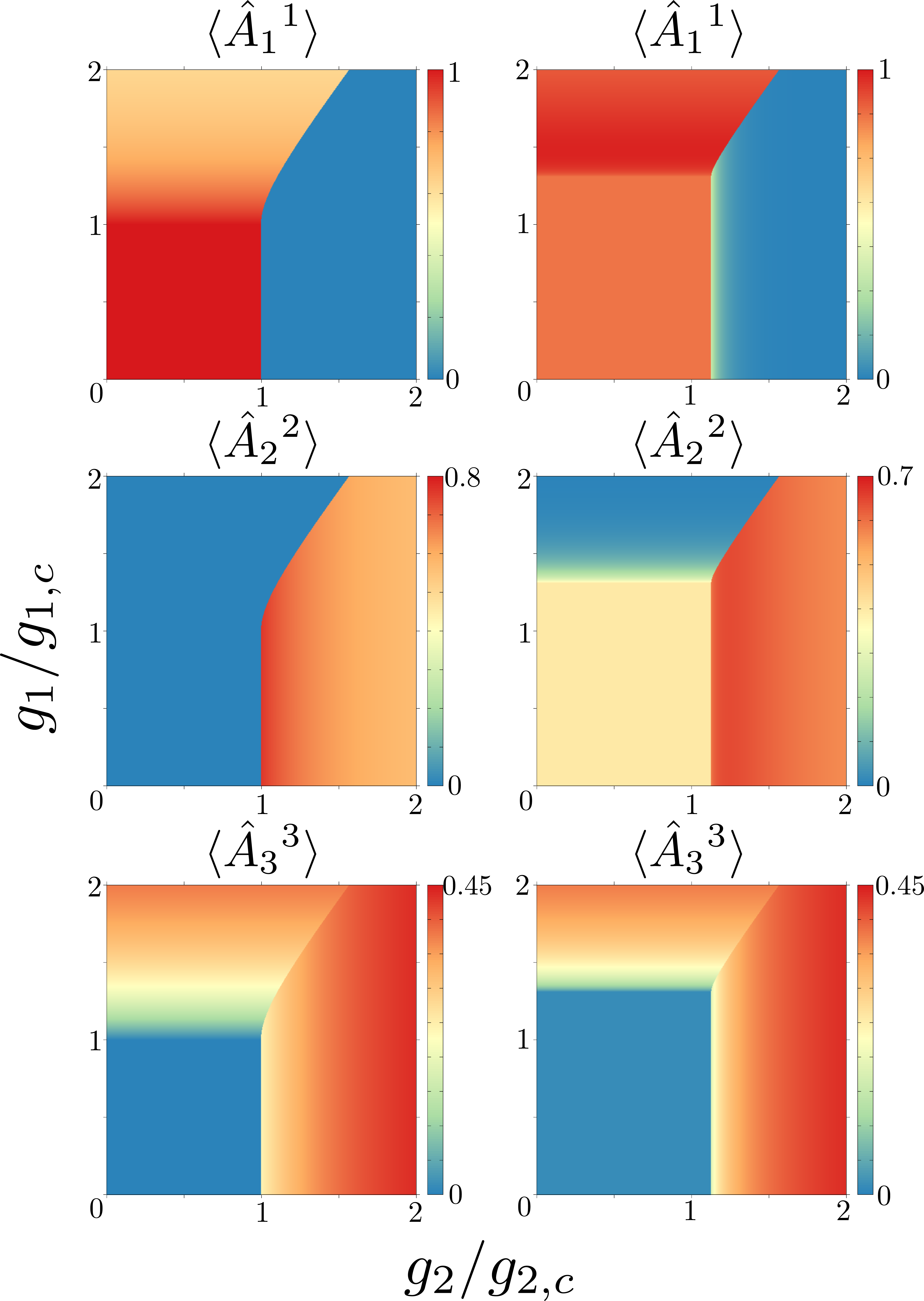}
	\caption{(Color online) Numerical computation of the scaled
    occupations $\braket{\Axy n n} \mathcal N$ of the single-particle
    energy levels of the three-level systems for $k_{\mathrm B} T =
    0.001 \, \Delta$ (left) and $k_{\mathrm B} T = 0.25 \, \Delta$
    (right). Parameters as in
    Fig.~\ref{fig:LM_ModeOccTs}.} \label{fig:LM_AtomOccTs}
\end{figure}
\begin{figure}[t]
  \centering
  \includegraphics[width=8.5cm]{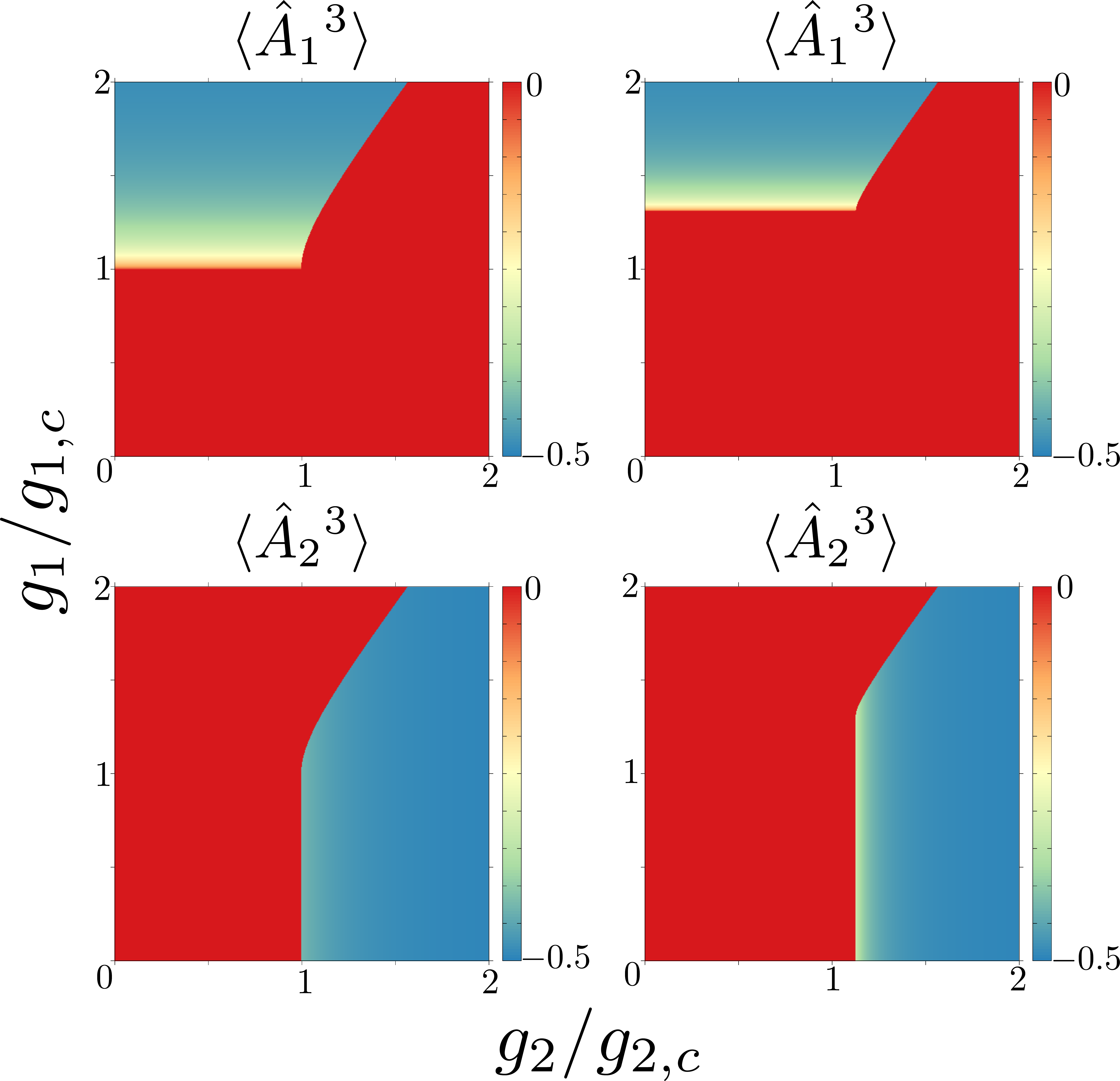}
	\caption{(Color online) Numerical computation of the scaled
    polarisations $\braket{\Axy 1 3} / \mathcal N$, $\braket{\Axy 2 3}
    / \mathcal N$ of the three-level systems for $k_{\mathrm B} T =
    0.001 \, \Delta$ (left) and $k_{\mathrm B} T = 0.25 \, \Delta$
    (right). Parameters as in
    Fig.~\ref{fig:LM_ModeOccTs}.} \label{fig:LM_AtomPolTs}
\end{figure}

Fig.~\ref{fig:LM_ModeOccTs}, \ref{fig:LM_AtomOccTs},
\ref{fig:LM_AtomPolTs} show the occupation $\braket{\hat a^\dag_n \hat
  a_n}$ of the modes of the resonator, the occupation $\Axy n n$ of
the single-particle levels of the atoms, and the polarisations $\Axy 1
3$, $\Axy 2 3$ of the atoms for low and high temperatures,
respectively. All plots have been generated numerically for finite
values of $\delta$.

These figures corroborate our findings from the analytical discussion
of the partition sum for vanishing $\delta$. We see three phases: a
\textit{normal} phase for coupling strengths $g_1$ and $g_2$ below the
critical coupling strengths $g_{1,c}$ and $g_{2,c}$, a \textit{red
superradiant} phase for large coupling strengths $g_1$ above the
critical coupling strength $g_{1,c}$, and a \textit{blue superradiant}
phase for coupling strengths $g_2$ above the critical coupling
strength $g_{2,c}$.

The normal phase is characterised by a zero occupation of both bosonic
modes (Fig.~\ref{fig:LM_ModeOccTs}) In addition, the polarisation, or
coherence, of the three-level systems is zero in the normal phase
(Fig.~\ref{fig:LM_AtomPolTs}).

In contrast, to the normal phase, the two superradiant phases are
characterised by a macroscopic occupation of only one of the two
bosonic modes; mode one in the red superradiant phase and mode two in
the blue superradiant phase. In addition, the red (blue) superradiant
phase shows a spontaneous polarisation $\braket{\Axy 1 3}$
($\braket{\Axy 2 3}$).

We see that these defining properties remain for increasing
temperature (right part of
Figs.~\ref{fig:LM_ModeOccTs}-\ref{fig:LM_AtomPolTs}). As discussed in
Sec.~\ref{sec:NormalPhasee}, we see that the population
$\braket{\Axy 2 2}$ of the single-particle energy level $\ket 2$
increases for rising temperature. The same is true for the occupation
$\braket{\Axy 3 3}$, though this is not visible in the right part of
Fig.~\ref{fig:LM_AtomOccTs} due to the fact that the temperature is
yet too small.

\begin{figure}[t]
  \centering
  \includegraphics[width=7.5cm]{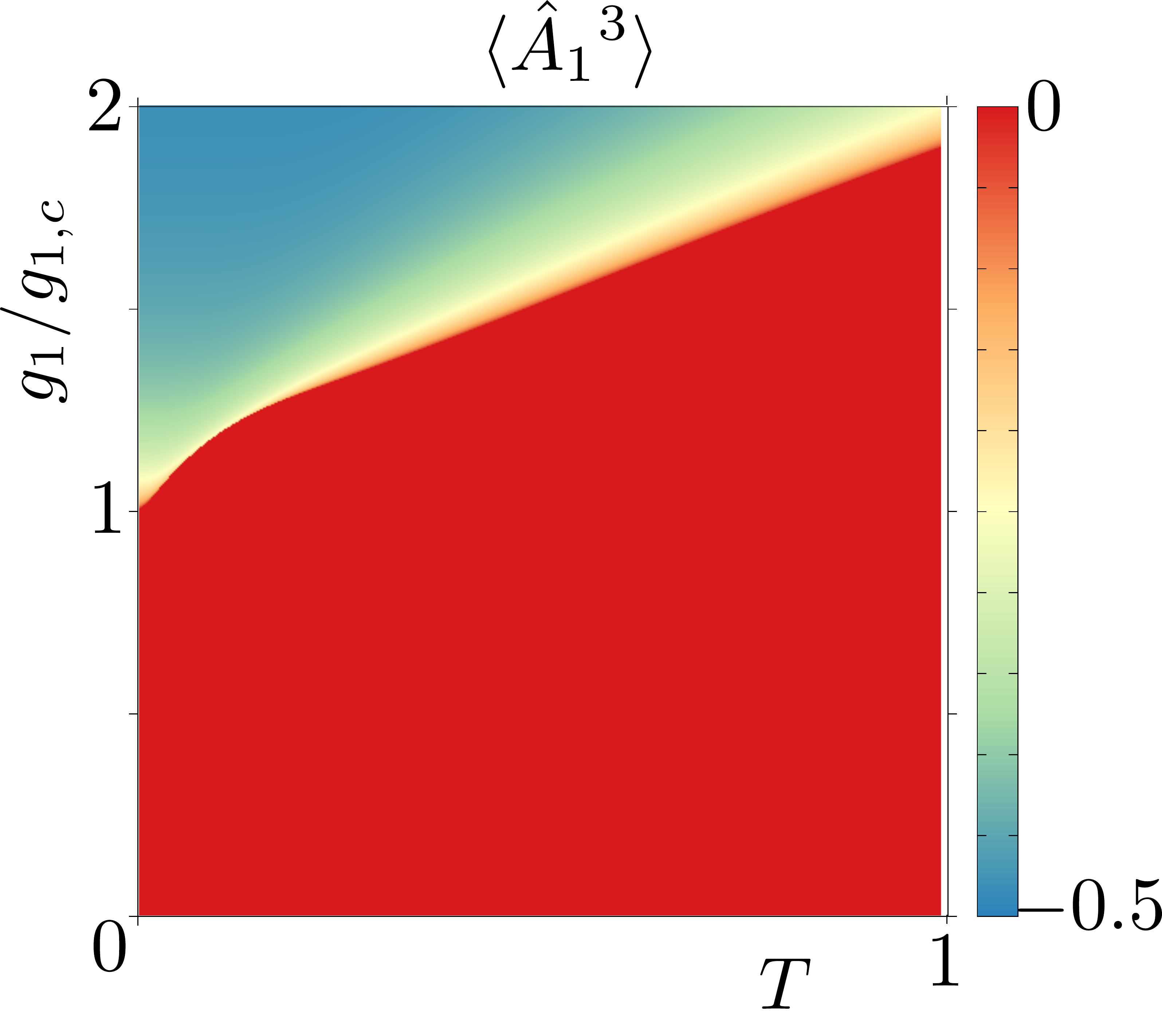}
	\caption{(Color online) Numerical computation of the scaled
    polarisation $\braket{\Axy 1 3} / \mathcal N$ of the transition
    $\ket 1 \leftrightarrow \ket 3$ of the three-level
    system. Parameters as in Fig.~\ref{fig:LM_ModeOccTs} and $g_2 =
    0.2 \, g_{2,c}$.} \label{fig:LM_A13_g1_T}
\end{figure}

From the Figs.~\ref{fig:LM_ModeOccTs}-\ref{fig:LM_AtomPolTs} we also
see that the shape of the phase boundary remains a straight line
between the normal and the two superradiant phases. Between the red
and the blue superradiant phases, the form of the phase boundary seems
to persist as well. The only effect of the rising temperature is a
shift of the phase boundary towards higher values of the coupling
strengths $g_1$ and $g_2$. This is visualised in
Fig.~\ref{fig:LM_A13_g1_T} where the polarisation $\braket{\Axy 1 3}$
of the transition $\ket 1 \leftrightarrow \ket 3$ of the three-level
systems is shown for variable coupling strength $g_1$ and temperature
$T$. The coupling strength of the second mode is fixed to $g_2 = 0.2
\, g_{2,c}$. We see that for increasing temperature, the superradiant
phase diminishes.

\begin{figure}[t]
  \centering
  \includegraphics[width=7.5cm]{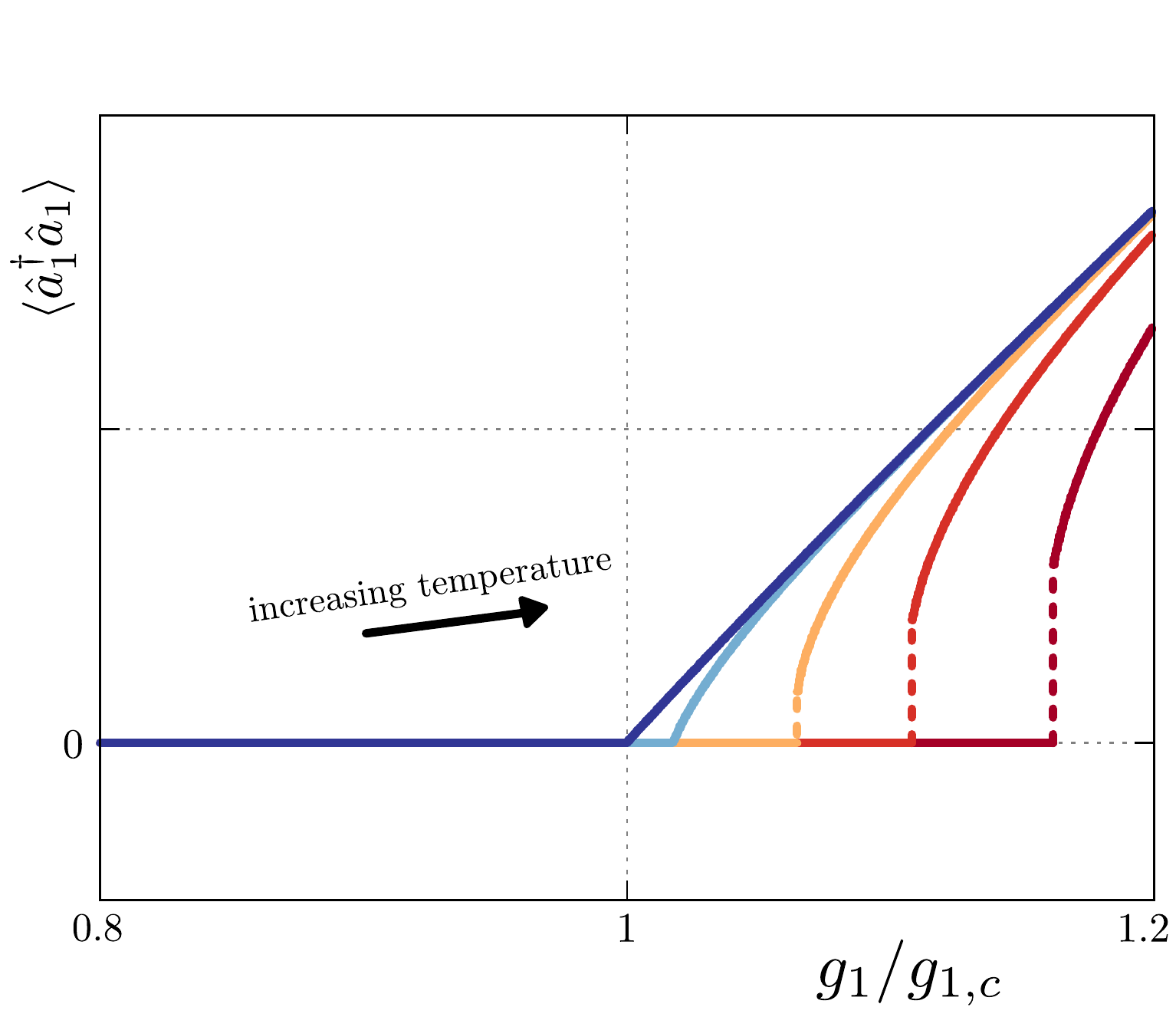}
	\caption{(Color online) Occupations $\braket{\hat a_1^\dag \hat
      a_1}$ of the first bosonic mode as a function of the coupling
    strength $g_1$ for fixed coupling strength $g_2 = 0.2 g_{2,c}$ and
    rising temperature (left to right, blue to red, respectively). The
    other parameters are set to $\Delta=1$, $\delta=0.1$,
    $\omega_1=1.1$, and $\omega_2=0.8$.} \label{fig:LM_n1jumps}
\end{figure}

In addition to the shift of the phase boundary, the jump in the
observables at this first-order phase transition increases. This is
shown in Fig.~\ref{fig:LM_n1jumps} for the occupation $\braket{\hat
  a_1^\dag \hat a_1}$ of the first bosonic mode. Of course,
numerically, jumps are hard to detect since we get a discrete set of
points as an output anyway. However, the dotted lines in
Fig.~\ref{fig:LM_n1jumps} connect two largely separated points; each
of the lines consists of 1000 data points. Thus we can really speak of
jumps in the observables and thus of a first-order phase transition.

\subsection{Zero-Temperature Limit}
Lastly, we analyse the zero-temperature limit of the
Lambda-model for $\delta = 0$. For decreasing temperature, the
function $q(\Omega)$, Eq.~\eqref{eq:LM_Def_gOmega}, becomes more and
more step function like. Indeed, for $\beta \Delta \gg 1$, $q(\Omega)$
can be written as a Fermi function, and eventually in the limit $\beta
\Delta \rightarrow \infty$, $q(\Omega)$ is given by
\begin{equation}
  q(\Omega) =
  \begin{cases}
    0 & , \quad \Omega < 1 \\
    1 & , \quad \Omega > 1.
  \end{cases}
\end{equation}
Hence at zero temperature, Eq.~\eqref{eq:LMFMinCondn} has always a unique
solution for coupling strengths $g_n > g_{n,c}$. In addition, since
$f(y)$ shows no additional maximum, we have a continuous phase
transition in this quantum limit.

Furthermore, we can also compute the position of the minimum of
$f$. In the limit $T \rightarrow 0$, Eq.~\eqref{eq:LMFMinCondn} reads
\begin{equation}
  0 = \Bigl( \frac{g_{n,c}}{g_n} \Bigr)^2 \sqrt{1 + \frac{4\hbar
      \omega_n}{\Delta} \Bigl( \frac{g_n}{g_{n,c}} \Bigr)^2 y_n^2 \;}
  - 1.
\end{equation}
Solving for $y_n$, we obtain
\begin{equation} \label{eq:LM_ZTL_yn}
  y_n = \pm \frac{g}{\hbar \omega_n} \sqrt{1 - \Bigl(\frac{g_{n,c}}{g_n}
    \Bigr)^4 \,}.
\end{equation}
If we identify $y_n$ with the mean-fields $\varphi_n$ of
Ref.~\cite{Hayn2011}, we reproduce our results \cite{Hayn2011}
for the superradiant phases of the quantum phase transition of the
Lambda-model.

The mean-fields $\Psi_n$ of Ref.~\cite{Hayn2011} can be reproduced as
well. Consider for instance $\Psi_3$ which is related to $\braket{\Axy
  3 3}$ through $\braket{\Axy 3 3} = \mathcal N \Psi_3^2$, except for
a possible phase. Using the results for the Boltzmann operator in the
red superradiant phase, Eq.~\eqref{eq:LMBSRExpOp}, plus the above
expression for $y_1$, Eq.~\eqref{eq:LM_ZTL_yn}, and finally insert
everything into the expectation value of Eq.~\eqref{eq:LM_ExpValAnm1},
we obtain
\begin{equation}
  \braket{\Axy 3 3} = \frac{\mathcal N}{2} \frac{1 - \mathrm e^{\beta \Delta
      \Omega} + \Bigl( 1 + \mathrm e^{\beta \Delta \Omega} \Bigr)
    \Omega}{\bigl[1 + \mathrm e^{\beta \Delta \Omega} + \mathrm
    e^{\frac{\beta \Delta}{2} (\Omega + 1)} \bigr] \Omega},
\end{equation}
which in the zero-temperature limit $\beta \Delta \rightarrow \infty$
reduces to
\begin{equation}
  \braket{\Axy 3 3} = \frac{\mathcal N}{2} \frac{\Omega - 1}{\Omega}.
\end{equation}
Here, $\Omega = \Omega(y_1)$ from Eq.~\eqref{eq:LM_Def_Omega3}. If we
finally insert the position $y_1$ of the minimum of the free energy,
Eq.~\eqref{eq:LM_ZTL_yn}, the population of the third single-particle
energy level in the zero-temperature limit is given by
\begin{equation}
  \braket{\Axy 3 3} = \frac{\mathcal N}{2} \Bigl[ 1 - \Bigl(
  \frac{g_{1,c}}{g_1} \Bigr)^2 \Bigr]
\end{equation}
which agrees with the findings of Ref.~\cite{Hayn2011} for
$\Psi_3$. Applying the same technique, we can obtain the expectation
values of all other collective atomic operators $\Anm$ in both
superradiant phases. This again coincides with the results of
Ref.~\cite{Hayn2011}.

\section{Conclusion}
We have analysed the Lambda-model in the thermodynamic limit at finite
temperatures using the partition sum of the Hamiltonian. Compared to
the quantum phase transition of this model~\cite{Hayn2011}, we found
that at finite temperatures the properties of the phases and phase
transition partially persist. Namely, we found three phases: a normal
and two superradiant phases. These have the same properties as in the
quantum limit. For small couplings and/or at high temperatures, the
system is in the normal phase where all particles are in their
respective single-particle ground-state and both bosonic modes are in
their vacuum state. If one of the coupling strengths is increased
above a temperature-dependent critical coupling strength, the system
undergoes a phase transition into a superradiant phase wich is
characterised by a macroscopic occupation of one of the bosonic modes
only and a spontaneaous polarisation of the corresponding branch of
the three-level system.

A new characteristic of the phase transition at finite temperatures is
the appearance of first-order phase transitions only. For the quantum
phase transition we already found first-order phase transitions
between the normal and the red superradiant phase and between the two
superradiant phases. Here, at finite temperatures, the phase
transition from the normal to the blue superradiant phase becomes a
first-order phase transition as well. This change of the order of the
phase transition would appear for a single bosonic mode as
well. Hence, it is due to the additional single-particle energy level
that first-order phase transitions show up. In addition, we emphasise
that even in the degenerate limit, $\delta \rightarrow 0 $, the phase
transitions from the normal to both superradiant phases are of first
order.

It is remarkable that in contrast to the original Dicke model, the
mean-field phase transitions at finite temperatures are not
continuous. This facet becomes significant if \textsl{real} atoms and
photons are considered. Here, for the original Dicke model with its
continuous phase transition, there exists a no-go theorem
\cite{Rzazewski1975, Rzazewski1976, Rzazewski1976a, Knight1978,
  Bialynicki-Birula1979, Slyusarev1979, Nataf2010,
  Viehmann2011}. However for first-order mean-field quantum phase
transitions, it is known \cite{Hayn2012, Baksic2013} that this no-go
theorem does not apply and superradiant phase transitions occur. We
expect an identical conclusion for the Lambda-model at finite
temperatures.

\section{Acknowledgements}
This work was supported by the Deutsche Forschungsgemeinschaft within
the SFB 910, the GRK 1558, and the projects BR 1528/82 and 1528/9.

\section{Appendix}
\subsection{Expectation Values of Bosonic Mode Operators}
\label{sec:AppA}

We calculate the expectation value of functions $G$ of operators of
the two bosonic modes as,
\begin{align} \label{eq:LM_ExpValModeOp}
  &\braket{G(\hat a^\dag_1, \hat a_1, \hat a^\dag_2, \hat a_2)} =
  \frac{1}{\mathcal Z} \mathrm{Tr} \{ G(\hat a^\dag_1, \hat a_1, \hat
  a^\dag_2, \hat a_2) \mathrm e^{-\beta \hat H}\} \\
  &= \frac{1}{\mathcal Z} \int_{\mathbb C^2} \frac{\mathrm d^2
    \alpha_1 \mathrm d^2\alpha_2}{\pi^2} G(\alpha_1^*, \alpha_1,
  \alpha_2^*, \alpha_2) \\
  &\qquad \times \mathrm e^{-\beta \sum_{n=1}^2 \hbar \omega_n \lvert
    \alpha_n \rvert^2} \mathrm{Tr} \{ \mathrm e^{-\beta \hat h}
  \}^{\mathcal N} \\
  &= \frac{1}{\mathcal Z} \frac{\mathcal N^2}{\pi^2} \int_{\mathbb
    R^2} \mathrm dy_1 \mathrm dy_2 G(\sqrt{\mathcal N} y_1,
  \sqrt{\mathcal N} y_1, \sqrt{\mathcal N} y_2, \sqrt{\mathcal N} y_2)
  \\
  &\qquad \times \mathrm e^{-\beta f} \\
 &= G(\sqrt{\mathcal N \,} y_{1,0}, \sqrt{\mathcal N \,} y_{1,0},
  \sqrt{\mathcal N \,} y_{2,0}, \sqrt{\mathcal N \,} y_{2,0}).
\end{align}
In the last step we used that the exponential dominates for large
$\mathcal N$. Hence, the function $G$ can be considered constant and
the remaining integral plus the prefactor is equal to the partition
sum.

\subsection{Expectation Values of Operators of the Three-Level
  Systems}
\label{sec:AppB}

For mean values of collective operators $\hat M$
(cf.~\ref{sec:LM_ExpValAtoms}) of the three-level systems we have,
\begin{align}
  \braket{\hat M} &= \frac{1}{\mathcal Z} \mathrm{Tr} \{ \hat M \mathrm
  e^{-\beta  \hat H} \} \\
  &= \frac{1}{\mathcal Z} \int_{\mathbb C^2} \frac{\mathrm d^2
    \alpha_1 \mathrm d^2 \alpha_2}{\pi^2} \mathrm e^{-\beta
    \sum_{n=1}^2 \hbar \omega_n \lvert \alpha_n \rvert^2}\\
  &\qquad \times \sum_{k=1}^{\mathcal N} \mathrm{Tr} \Bigl\{ \hat m^{(k)}
  \mathrm e^{-\beta \hat h^{(1)}} \cdot \ldots \cdot \mathrm e^{-\beta
  \hat h^{(\mathcal N)}} \Bigr\} \\
  &= \frac{1}{\mathcal Z} \int_{\mathbb C^2} \frac{\mathrm d^2
    \alpha_1 \mathrm d^2 \alpha_2}{\pi^2} \mathrm e^{-\beta
    \sum_{n=1}^2 \hbar \omega_n \lvert \alpha_n \rvert^2} \\
  &\qquad \times \mathrm{Tr} \{ \mathrm e^{-\beta \hat h}\}^{\mathcal N}
  \mathcal N \frac{\mathrm{Tr} \Bigl\{ \hat m \mathrm e^{-\beta \hat
      h} \Bigr\}}{\mathrm{Tr} \{ \mathrm e^{-\beta \hat h}\}} \\
  &= \mathcal N \frac{\mathrm{Tr} \Bigl\{ \hat m \, \mathrm e^{-\beta
      \hat h_0} \Bigr\}}{\mathrm{Tr} \{ \mathrm e^{-\beta \hat
      h_0}\}} \\
  &\equiv \mathcal N \braket{\hat m}_0,
\end{align}
with $\hat h_0 = \hat h(y_{1,0}, y_{2,0})$.

%
%

%
%

\end{document}